\documentclass[10pt]{iopart}

\usepackage{graphicx}

\usepackage{verbatim}
\usepackage[dvips]{color}  
\usepackage{ulem} 

\begin{document}

\title[Large anisotropy in the overdoped region of Ba(Fe$_{1-x}$Ni$_x$)$_2$As$_2$]{Large increase of the anisotropy factor in the overdoped region of Ba(Fe$_{1-x}$Ni$_x$)$_2$As$_2$ as probed by fluctuation spectroscopy}

\author{A Ramos-\'Alvarez$^1$, J Mosqueira$^1$, F Vidal$^1$, Xingye Lu$^2$, Huiqian Luo$^2$}

\address{$^1$LBTS, Facultade de F\'isica, Universidade de Santiago de Compostela, E-15782 Santiago de Compostela, Spain}

\address{$^2$Beijing National Laboratory for Condensed Matter Physics, Institute of Physics, Chinese Academy of Sciences, Beijing 100190, China}

\ead{j.mosqueira@usc.es}

\begin{abstract}
We study the diamagnetism induced by thermal fluctuations above the superconducting transition of the iron pnictide Ba(Fe$_{1-x}$Ni$_x$)$_2$As$_2$ with different doping levels. The measurements are performed with magnetic fields up to 7~T applied in the two main crystal directions. These data provide double information: first, they confirm at a quantitative level the applicability to these materials of a 3D-anisotropic Ginzburg-Landau approach valid in the finite field regime. Then, they allow to determine the doping-level dependence of the in-plane coherence length and of the superconducting anisotropy factor, $\gamma$. Our results provide a stringent confirmation of the large increase of $\gamma$ with the doping level, as recently proposed from magnetoresistivity measurements. The implications of the applicability of the model used to a multiband superconductor are discussed.
\end{abstract}

\submitto{\SUST}
\maketitle

\ioptwocol

\section{Introduction}

In addition to its intrinsic interest, thermal fluctuation near a superconducting transition are a very useful tool (sometimes named fluctuation spectroscopy) to obtain fundamental superconducting parameters like the upper critical field, the superconducting coherence lengths, the anisotropy, and even the effective dimensionality of the material under study \cite{tinkham}. The usefulness of this technique becomes evident in high temperature superconductors (cuprates or Fe-based). In these materials, commonly studied observables (resistivity, magnetic susceptibility, specific heat, etc) very often present a rounded behavior around the transition temperature, $T_c$, due precisely to superconducting fluctuations \cite{reviewsFeSC,tinkham2}. This rounding, which is enhanced by the application of a magnetic field, $H$, complicates the direct determination of the above mentioned central parameters, and a comparison with existing models for the effect of superconducting fluctuations is required to analyze the experimental data.

The fluctuation effects have been recently used in a number of works to characterize the superconducting properties of Fe-based superconductors, through observables like the electrical conductivity, magnetization, or specific heat \cite{pallecchi,fanfarillo,salemsugui,choi,putti,liuPLA10,tesanovic,pandyaSST10,kim,liuSSC11,pandyaSST11,mosqueira,welpPRB11,liu2,prando,song,marra,rullier,reySST13,salemsuguiSST13,mosqueiraJSNM13,reySST14}. Nowadays, the properties of these materials at the optimal doping were extensively studied and are already rather well understood. However, their behavior at doping levels far from the optimal one are much less investigated and some aspects remain still open. An important example is the large anisotropy observed in Ba(Fe$_{1-x}$Ni$_x$)$_2$As$_2$ as follows from recent measurements of the fluctuation in-plane magnetoconductivity: in this compound the anisotropy factor (defined as the ratio between the in-plane and transverse coherence lengths) increases from $\gamma=2$ at optimal doping ($x=0.05$) up to around $\gamma=15$ for $x=0.10$ \cite{reySST13}. To the best of our knowledge, such a $\gamma$ value is the largest reported for an iron pnictide of the 122 family, and it is even larger than the one observed in some high-$T_c$ cuprates (e.g., optimally doped YBa$_2$Cu$_3$O$_{7-\delta}$). It is worth noting, however, that these results may be affected by large uncertainties (up to 30\%) in the geometry of the crystals and of the electrical contacts and, therefore, further verification is desirable.

Here we present measurements of the fluctuation-induced magnetic susceptibility above $T_c$, $\chi_{\rm fl}$, in Ba(Fe$_{1-x}$Ni$_x$)$_2$As$_2$ with different doping levels. The data were taken with magnetic fields up to 7~T applied in the two main crystallographic directions (parallel and perpendicular to the FeAs ($ab$) layers). The interest of these measurements is twofold. On the one side, previous works on the diamagnetism induced by superconducting fluctuations mainly focus in the critical region close to the $T_c(H)$ line. Only one work (about optimally doped Ba$_{1-x}$K$_x$Fe$_2$As$_2$) address the Gaussian region above $T_c(H)$ \cite{mosqueira}. Thus, the present work would allow to check the applicability of the phenomenological 3D-anisotropic Ginzburg-Landau (3D-aGL) approach to describe $\chi_{\rm fl}$ in this region as a function of the doping level. But also of particular interest is the behavior at high reduced magnetic fields, at which multiband effects could be observable \cite{koshelev14}. On the other side, $\chi_{\rm fl}$ is highly dependent on the orientation of the applied magnetic field, and it will be very useful to accede experimentally to the anisotropy factor. In particular, the 3D-aGL approach predicts that in the zero-field limit $\chi^\perp_{\rm fl}/\chi_{\rm fl}^\parallel\approx\gamma^2$, where the superscripts $\perp$ and $\parallel$ correspond to $H\perp ab$ and $H\parallel ab$, respectively. Thus, the simultaneous measurement of both $\chi_{\rm fl}^\perp$ and $\chi_{\rm fl}^\parallel$ would allow to confirm the striking increase of $\gamma$ upon overdoping observed in Ref.~\cite{reySST13}. 

\section{Experimental details and results}

\subsection{Crystals fabrication and characterization}

The Ba(Fe$_{1-x}$Ni$_x$)$_2$As$_2$ samples used in this work are plate-like single crystals (see Table~1) with the crystal $ab$ layers parallel to the largest faces. They were cleaved from larger crystals grown by the self-flux method. Their nominal Ni doping levels are $x=0.05$, 0.075, 0.09 and 0.10, although the actual doping level was found to be a factor $\sim0.8$ smaller (see Ref.~\cite{growth}, where all the details of the growth procedure and characterization may be found).

\subsection{Superconducting transition temperatures and transition widths}

The magnetization measurements were performed with a commercial SQUID magnetometer (Quantum Design, model MPMS-XL) with magnetic fields up to 7~T. As commented above, the measurements were performed with both $H\parallel ab$ and $H\perp ab$. In the first case the crystals were glued with GE varnish to a quartz sample holder (0.3~cm in diameter, 22~cm in length) with two plastic rods at the ends which ensured an alignment better than $0.1^\circ$. For the measurements with $H\perp ab$ we made a groove ($\sim0.3$ mm wide) in the sample holder into which the crystals were glued also with GE varnish. The crystal alignment was checked by optical microscopy to be better than $5^\circ$. This allowed to determine the anisotropy factor from the anisotropy of the precursor diamagnetism with a $\sim0.5$\% uncertainty.\footnote{According to the 3D-aGL approach in the low-field limit, if the crystal misalignment when measuring with $H\perp ab$ ($H\parallel ab$) is $\theta_\perp$ ($\theta_\parallel$), the measured $\chi_{\rm fl}^\perp/\chi^\parallel_{\rm fl}$ would be given by $\gamma_{\rm eff}^2=(\gamma^2\cos^2\theta_\perp+\sin^2\theta_\perp)/(\gamma^2\sin^2\theta_\parallel+\cos^2\theta_\parallel)$ (see, e.g., Ref.~\cite{alignment}). By using $\theta_\perp=5^\circ$, $\theta_\parallel=0.1^\circ$, and $\gamma\sim 2-15$ (see Ref.~\cite{reySST13}), $\gamma_{\rm eff}$ would be within 0.5\% the actual $\gamma$ value.}

\begin{figure}[t]
\begin{center}
\includegraphics[scale=.5]{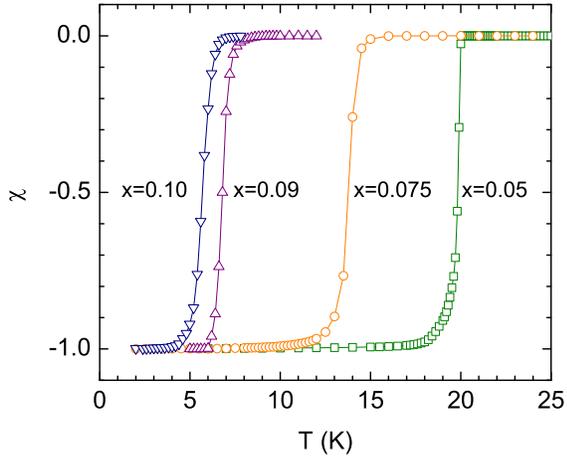}
\caption{Temperature dependence of the ZFC magnetic susceptibility of the samples studied (already corrected for demagnetizing effects) obtained with a $0.5$ mT perpendicular to the $ab$-layers. The $x$ value represents the doping level.}
\label{figure1}
\end{center}
\end{figure}

In Fig.~\ref{figure1} it is presented the temperature dependence of the zero-field-cooled (ZFC) magnetic susceptibility for all crystals studied, measured with a $0.5$ mT field applied perpendicular to the $ab$ layers. The demagnetizing effect was corrected by using the demagnetizing factors $D$ needed to attain the ideal value of -1 at low temperatures, which are within 5\% the ones resulting from the crystals shape (see Table 1). From these curves, $T_c$ was estimated by linearly extrapolating to $\chi=0$ the higher-slope $\chi(T)$ data, and the transition width as $\Delta T_c=T_{c0}-T_c$, where $T_{c0}$ is the highest temperature at which a diamagnetic signal is resolved in these low-field measurements. The results are also compiled in Table 1. As expected \cite{reviewsFeSC}, $T_c$ is reduced upon doping above the optimal doping level. The observed $T_c(x)$ dependence is consistent with the one found in Ref.~\cite{niPRB10} in Ba(Fe$_{1-x}$Ni$_x$)$_2$As$_2$ from measurements in different observables. The transition widths increase from $\sim0.3$~K in the optimally doped compound, to $\sim0.6$~K in the overdoped compounds. These values are very close to the ones determined from the resistive transition in crystals from the same batches, see Ref.~\cite{reySST13}. In relative terms, the $\Delta T_c/T_c$ increase is significant (from $\sim0.015$ to $\sim0.1$ on increasing $x$ from 0.05 to 0.10). A similar effect is also present in high-$T_c$ cuprates \cite{intrinsic}, and may be due to $T_c$ inhomogeneities intrinsic to the non-stoichiometric nature of these compounds. In any case, as we will see below, these $\Delta T_c$ values will allow to study fluctuation effects in a wide temperature region above $T_c(H)$. 

\begin{table}[t]
\begin{center}
\begin{tabular}{cccccc}
\hline
$x$  & $L_a\times L_b\times L_c$ & $D$ & $T_c$ & $\Delta T_c$ & $T_{onset}$\\ 
        &  (mm$^3$) &  & (K) & (K) & (K)\\
\hline  
0.05 & $3.7\times1.45\times0.32$ & 0.77 & 20.0 & 0.3 & 27.0\\  
0.075 & $1.5\times1.20\times0.13$ & 0.82 & 14.2 &  0.6 & 18.2\\ 
0.09 & $4.2\times1.85\times0.07$ & 0.90 & 7.2 & 0.8 & 10.2 \\ 
0.10 & $2.0\times1.80\times0.10$ & 0.92 & 6.3 & 0.6  & 8.9 \\ 
\hline
\end{tabular}
\end{center}
\caption{Some parameters of the crystals studied relevant for the analysis. See the main text for details.}
\label{tabla1}
\end{table}

\subsection{Fluctuation contribution to the magnetic susceptibility above $T_c$}

To measure the weak magnetic moment due to superconducting fluctuations above $T_c$ ($m\sim -10^{-5}$ emu in the samples used) we used the \textit{Reciprocating Sample Option} (RSO). We averaged eight measurements consisting of 10 cycles at 1 Hz frequency, which lead to a resolution in the $\sim10^{-8}$ emu range. In the present experiments we have used magnetic fields $\mu_0H\geq3$~T. This allowed us to analyze the data with the conventional GL approach described below (at lower field amplitudes it has been reported that the fluctuation effects in these materials are strongly enhanced with respect to conventional GL approaches, due to the possible presence of phase fluctuations \cite{prando} and/or $T_c$ inhomogeneities \cite{reySST13}). 

\begin{figure*}[t]
\begin{center}
\includegraphics[scale=.7]{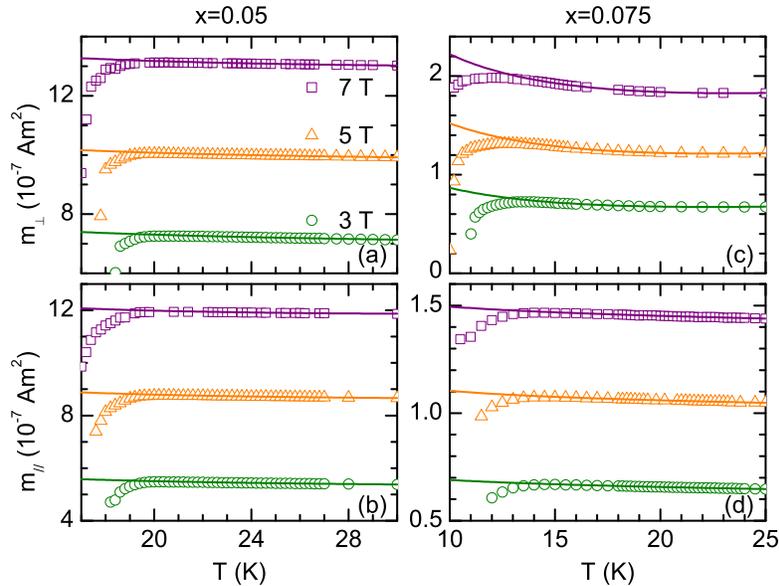}
\caption{Example, for two of the studied samples, of the temperature dependence of the magnetic moment above $T_c$. Upper (lower) panels were obtained with $H\perp ab$ ($H\parallel ab$). The normal-state backgrounds (lines) were determined by fitting a Curie-like function (Eq.~(\ref{back})) above $\sim1.3T_c$, where fluctuation effects are negligible. For details see the main text.}
\label{figure2}
\end{center}
\end{figure*}

Some examples of the as-measured $m(T)$ data around $T_c$ are presented in Fig.~\ref{figure2}, where the  rounding associated to superconducting fluctuations may already be appreciated. For each applied field, the temperature dependence of the fluctuation magnetic moment was obtained through
\begin{equation}
m_{\rm fl}(T)=m(T)-m_B(T)
\end{equation}
where $m_B(T)$ is the background contribution due to the samples normal state and to the sample holder. This last was determined by fitting a Curie-like function
\begin{equation}
m_B(T)=c_1+c_2T+\frac{c_3}{T}
\label{back}
\end{equation}
to the raw data in a temperature interval from $\sim 1.3 T_c$ up to above $\sim1.8T_c$ ($c_1$, $c_2$ and $c_3$ are free parameters). The lower bound of this fitting region corresponds to a reduced temperature $\varepsilon\equiv\ln(T/T_c)\approx0.3$, above which fluctuation effects in these materials are expected to be negligible \cite{reySST13,reySST14}. The resulting $m_B(T)$ contributions are presented as solid lines in Fig.~\ref{figure2}. 

The resulting fluctuation magnetic susceptibility, $\chi_{\rm fl}(T)=m_{\rm fl}(T)/HV$ (where $V$ is the crystals volume estimated from their mass and from the theoretical density), is presented in Fig.~3 for all studied samples and for both $H\perp ab$ and $H\parallel ab$. Some qualitative aspects may be directly obtained from this figure:
i) The rounded $\chi_{\rm fl}(T)$ behavior extends several Kelvin above $T_c$ for all doping levels, up to an onset temperature $T_{onset}\approx1.3T_c$ (see Table 1) which is well beyond the corresponding transition widths. This indicates that $T_c$ inhomogeneities may play a negligible role. 
ii) The fluctuation magnetic susceptibility is anisotropic, being significantly larger in amplitude when $H\perp ab$. This anisotropy increases appreciably with the doping level, which is already consistent with the large increase of the anisotropy factor observed in Ref.~\cite{reySST13} in the same compounds above the optimal doping.
iii) The $\chi_{\rm fl}$ amplitude decreases with the magnetic field (mainly when it is applied perpendicular to the $ab$ layers, due to the anisotropy of the upper critical field). This indicates that the fields used in the experiments are large enough as to enter in the finite field (or Prange) fluctuation regime, where $\chi_{\rm fl}$ strongly decreases with $H$ \cite{soto04}. The quantitative analysis of the data would then require using theoretical approaches valid beyond the zero-field (or Schmidt) limit.

\begin{figure*}[t]
\begin{center}
\includegraphics[scale=.6]{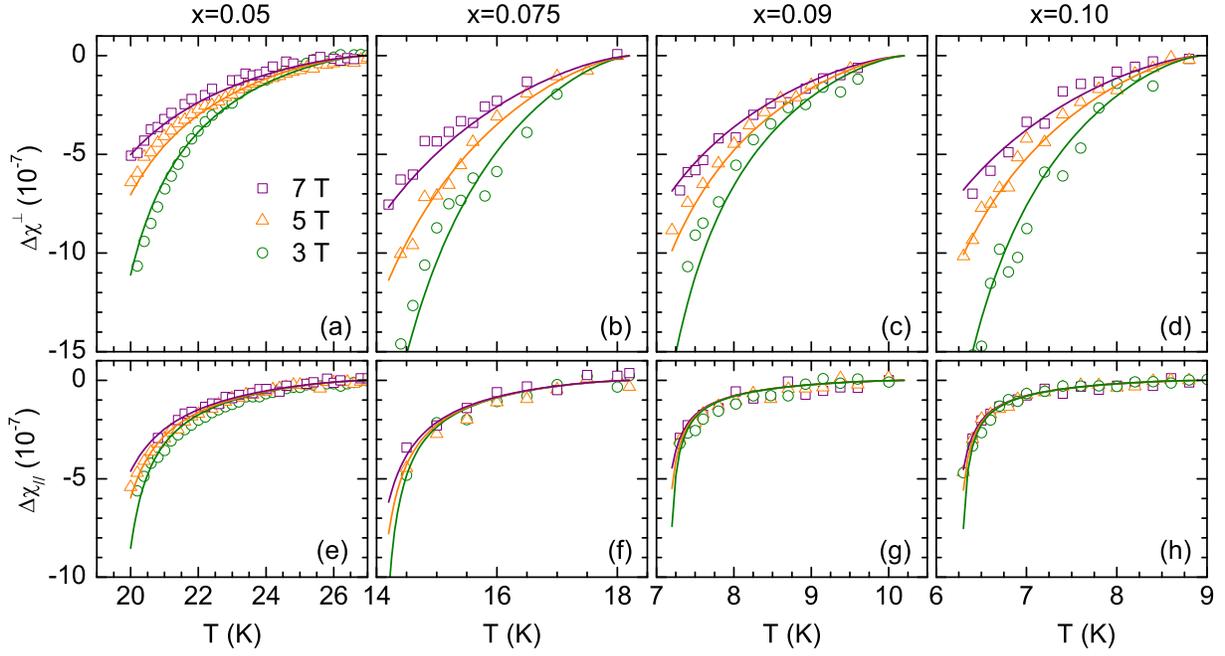}
\caption{Temperature dependence just above $T_c$ of the fluctuation magnetic susceptibility for all studied doping levels. Upper (lower) panels correspond to $H\perp ab$ ($H\parallel ab$). The lines in the upper panels are the best fits of the 3D-aGL approach for $H\perp ab$ (Eq.~(\ref{prange})) with $\xi_{ab}(0)$ and $\xi_c(0)$ as the only free parameters for each doping level. The lines in the lower panels were obtained \textit{without free parameters}, by using in the 3D-aGL expression for $H\parallel ab$ (Eq.~(\ref{prangepara})) the same coherence lengths.}
\label{figure3}
\end{center}
\end{figure*}

\section{Theoretical background}

In spite of the multiband nature of the compound under study, previous measurements of the fluctuation-induced conductivity and magnetoconductivity were successfully explained in terms of a GL approach for single-band three-dimensional anisotropic supercondutors (3D-aGL approach), and it will be our starting point. Below we will comment on the implications of the applicability of single-band approaches to these materials.  

\begin{figure}[t]
\begin{center}
\includegraphics[scale=.5]{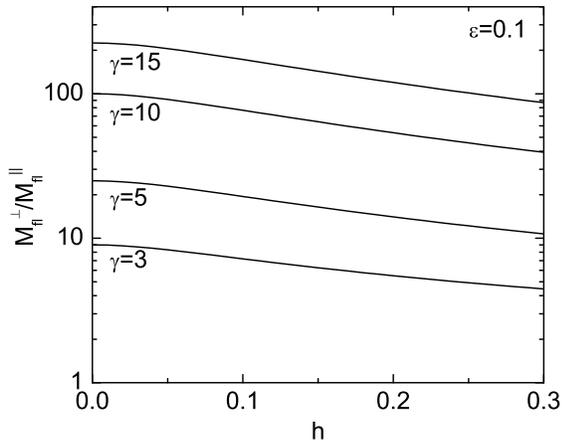}
\caption{Effect of a finite applied magnetic field on the $M_{\rm fl}^\perp/M_{\rm fl}^\parallel$ ratio, according to Eqs.~(\ref{prange}) and (\ref{prangepara}). While it is $\gamma^2$ when $h\to0$, it is slightly reduced on increasing $h$.}
\label{figure4}
\end{center}
\end{figure}

In terms of the 3D-aGL approach the fluctuation magnetization $M_{\rm fl}$ of an anisotropic superconductor (in presence of a field applied in the two main crystallographic directions) may be related to that of an isotropic superconductor through \cite{Klemm80, Blatter92, Hao92}
\begin{equation}
M^{\perp}_{\rm fl}(T,H)=\gamma M^{\rm iso}_{\rm fl}(T,H)
\label{aniperp}
\end{equation}
for $H\perp ab$, and 
\begin{equation}
M^{\parallel}_{\rm fl}(T,H)=M^{\rm iso}_{\rm fl}(T,H/\gamma)
\label{anipara}
\end{equation}
for $H\parallel ab$. In the low-field limit, i.e. for $H\ll H_{c2}(0)$, $M^{\rm iso}_{\rm fl}$ is given by Schmidt's classic result \cite{schmidt},
\begin{equation}
M^{\rm iso}_{\rm fl}(T,H)=-\frac{\pi k_BT\mu_0H\xi(0)}{6\phi_{0}^2}\varepsilon^{-1/2},
\label{schmidt}
\end{equation}
where $k_B$ is the Boltzmann constant, $\mu_0$ is the vacuum magnetic permeability, $\phi_0$ is the magnetic flux quantum, $\xi(0)$ is the coherence length, and $\varepsilon=\ln (T/T_c)$ is the reduced temperature. In this case, as $M_{\rm fl}^{\rm iso}\propto H$, the anisotropy factor could be obtained directly from the ratio
\begin{equation}
\frac{M_{\rm fl}^{\perp}(T,H)}{M_{\rm fl}^{\parallel}(T,H)}=\gamma^2.
\label{gamma2}
\end{equation}
As commented in \S2.3, the $H$ amplitudes used in the present experiments are beyond the low-field limit and Eqs.~(\ref{schmidt}) and (\ref{gamma2}) are not directly applicable. Buzdin and Feinberg derived an expression for the fluctuation magnetization of 3D anisotropic materials valid for arbitrary field amplitudes and orientations \cite{Buzdin}. However, their approach do not take into account short-wavelength effects, that may be relevant at high reduced temperatures or magnetic fields ($\varepsilon$ or $h$ of the order of 1). Here $h\equiv H/H_{c2}(0)$, where $H_{c2}(0)$ is the upper critical field for $H\perp ab$ linearly extrapolated to 0 K. In Refs.~\cite{mosqueiraPRL00,carballeira3D} it was shown that the introduction of a \textit{total-energy} cutoff in the fluctuation spectrum extends the applicability of the GGL approach to these short wavelength regimes. By combining the expression for $M_{\rm iso}$ in Ref.~\cite{carballeira3D} with Eqs.~(\ref{aniperp}) and (\ref{anipara}) it is obtained
\begin{eqnarray}
&&M^{\perp}_{\rm fl}(T,H)=-\frac{k_BT\gamma}{\pi\phi_0\xi_{ab}(0)}\int_{0}^{\sqrt{c-\varepsilon}}dq\left[\frac{c-\varepsilon}{2h}\right.\nonumber\\
&-&\ln\Gamma\left(\frac{\varepsilon+h+q^2}{2h}\right)+\left(\frac{\varepsilon+q^2}{2h}\right)\psi\left(\frac{\varepsilon+h+q^2}{2h}\right)\nonumber\\
&+&\left.\ln\Gamma\left(\frac{c+h+q^2}{2h}\right)-\left(\frac{c+q^2}{2h}\right)\psi\left(\frac{c+h+q^2}{2h}\right)\right]
\label{prange}
\end{eqnarray}
and 
\begin{equation}
M_{\rm fl}^{\parallel}(T,H)=\frac{1}{\gamma}M_{\rm fl}^{\perp}(T,H/\gamma).
\label{prangepara}
\end{equation}
Here $\Gamma$ and $\Psi$ are, respectively, the gamma and digamma functions, $\xi_{ab}(0)$ is the in-plane coherence length, and $c$ is a cutoff constant which value is expected to be close to 0.5 \cite{EPLcutoff}. In the low magnetic field limit ($h\ll\varepsilon$), in absence of cutoff ($c\to\infty$), and for isotropic materials ($\gamma=1$), Eq.~(\ref{prange}) reduces to  the Schmidt result, Eq.~(\ref{schmidt}). Equation (\ref{prange}) was successfully used to explain $\chi_{\rm fl}$ above $T_c$ in compounds like MgB$_2$ and NbSe$_2$ \cite{MgB2,NbSe2}, while their 2D and 2D-3D analogs accounted for the behavior of cuprate high-$T_c$ compounds \cite{intrinsic,YBCO,Tl2223,TlPb1212}. It is worth noting that, in view of Eqs.~(\ref{prange}) and (\ref{prangepara}), the ratio $M_{\rm fl}^\perp/M_{\rm fl}^\parallel$ decreases with $h$ below $\gamma^2$. However, as it is illustrated in Fig.~\ref{figure4}, for the $h$ values used in the experiments, and the $\gamma$ values expected after Ref.~\cite{reySST13}, such a ratio will be large enough as to determine $\gamma$ with a good accuracy. 

\section{Analysis and discussion}

\subsection{Comparison with the theory}

In order to compare the present measurements with the theory, we first fitted the Eq.~(\ref{prange}) (normalized by the applied field) to the $\chi_{\rm fl}^\perp(T,H)$ data in Fig.~\ref{figure3}. The fitting region range from $T_c(H=0)$ up to 1.3$T_c(H=0)$. The lower bound was chosen to avoid entering into the so-called critical region, where the Gaussian approximation is no longer valid.\footnote{According to the field-dependent Ginzburg criterion \cite{Ikeda} the upper bound of the critical region is given by $T_c^\perp(H)+T_c[4\pi k_B\mu_0H\gamma/\Delta c\xi_{ab}(0)\phi_0]^{2/3}$ when $H\perp ab$, and by $T_c^\parallel(H)+T_c[4\pi k_B\mu_0H/\Delta c\xi_{ab}(0)\phi_0]^{2/3}$ when $H\parallel ab$ ($\Delta c$ is the specific heat jump at $T_c$) \cite{mosqueira}. Recent measurements in a 122 iron pnictide with similar superconducting parameters (Ba$_{1-x}$K$_x$Fe$_2$As$_2$) revealed that, for the $H$ amplitudes used in our experiments, the upper bound of the critical region may be approximated by $T_c(H=0)$ for both field orientations \cite{mosqueira}.} In turn, the upper bound corresponds to the temperature above which fluctuation effects vanish, $T_{onset}\approx1.3T_c$ (see Table 1). As the cutoff constant corresponds to the reduced temperature for the onset of fluctuation effects, we will use $c=\ln(T_{onset}/T_c)\approx0.3$, a value consistent with the one found in previous works in the same material \cite{reySST13,reySST14}. For each doping level the only free parameters are $\xi_{ab}(0)$ and $\gamma$. Note that $\xi_{ab}(0)$ is present in the prefactor of Eq.~(\ref{prange}) but also in the reduced magnetic field, that may be expressed as $h=2\pi\mu_0H\xi_{ab}^2(0)/\phi_0$. As it may be seen in Figs.~\ref{figure3} (a) to (d), the agreement with the data is excellent for all field amplitudes and for all doping levels.
The fluctuation magnetization for $H\parallel ab$, Eq.~(\ref{prangepara}), depends on the same superconducting parameters. Therefore, the lines in Figs.~\ref{figure3} (e) to (h) were obtained \textit{without free parameters}, by just using in Eq.~(\ref{prangepara}) the $\xi_{ab}(0)$ and $\gamma$ values previously determined. As it may be seen, the agreement with the data is also excellent, strongly supporting the reliability of the resulting values for $\xi_{ab}(0)$ and $\gamma$, and the applicability of the single-band 3D-aGL approach used. It has been proposed that multiband effects may be observable when there is a large difference between the coherence lengths in different bands \cite{koshelev14}. In this case, a deviation from single-band approaches should appear at the field scale associated to the larger coherence length, $\xi_1$, i.e., $H_1=\phi_0/2\pi\mu_0\xi_1^2$. Our results could suggest that in this material the coherence lengths in different bands are not too different, or that $H_1$ is not in the field range explored in the present work. In fact, $H_1$ is the field scale at which the upper critical field has upward curvature \cite{koshelev14}, and recent magnetotransport measurements in the same compounds revealed that the upper critical field presents a linear $T$-dependence up to well above the $H$ values used here \cite{salemsuguiSST13,reySST14}. 

\begin{figure}[h]
\begin{center}
\includegraphics[scale=.5]{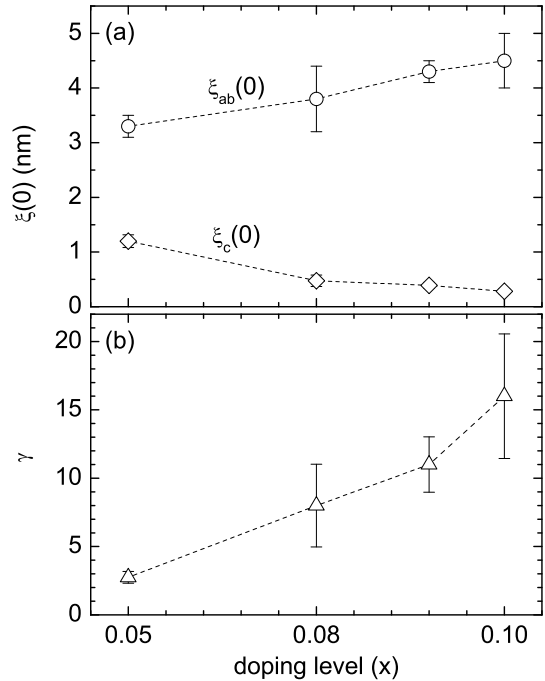}
\caption{Dependence with the doping level of the coherence length amplitudes (a) and of the anisotropy factor (b), as follows from the comparison of Eqs.~(\ref{prange}) and (\ref{prangepara}) with the data in Fig.~\ref{figure3}.}
\label{figure5}
\end{center}
\end{figure}

\subsection{Dependence of the superconducting parameters on the doping level}

In Fig.~\ref{figure5} we present the $\xi_{ab}(0)$ and $\gamma$ values resulting from the above analysis as a function of the doping level. This figure also includes the transverse coherence length amplitude, obtained as $\xi_c(0)=\xi_{ab}(0)/\gamma$. As it may be seen, $\xi_{ab}(0)$ increases moderately from 3.3~nm at optimal doping ($x=0.05$) to 4.5 nm well inside the overdoped region ($x=0.10$). However, $\gamma$ presents a pronounced increase from $\sim3$ to $\sim16$ in the same interval of doping levels. We are not aware of theoretical studies about a possible dependence of the anisotropy factor with the doping level in these compounds. However, a density functional study by Singh and Du \cite{singh} in LaFeAsO$_{1-x}$F$_x$ shows that the Fermi-surface sheets and dimensionality strongly depend on the doping level, and that the anisotropy tends to increases when the system is doped away from the parent phase. 

Associated to the increase in $\gamma$, there is a significant decrease of the transverse coherence length, from $\xi_c(0)\simeq1.2$~nm at optimal doping to $\xi_c(0)\simeq0.4$~nm for $x=0.1$. This last value is even smaller than the FeAs layers interdistance, $s\simeq0.64$~nm, which could suggest the possible presence of two-dimensional (2D) fluctuation effects. However, according to the Lawrence-Doniach model for a system of Josephson-coupled superconducting layers, the 2D behavior is expected to occur for reduced temperatures above $\sim(2\xi_c(0)/s)^2$, which is as large as 0.77 even for the most anisotropic crystal ($x=0.1$). This is well above the reduced temperature at which fluctuation effects are observe to vanish ($\varepsilon\simeq0.3$), confirming the adequacy of the 3D approach used to analyze the data.

\subsection{Comparison with $\gamma$ values in the literature for Fe-based superconductors}

The $x$-dependence of the superconducting parameters presented in Fig.~\ref{figure5} is consistent with the one obtained in Ref.~\cite{reySST13} from the fluctuation-induced in-plane magnetoconductivity of single crystals of the same composition, including the large increase of the anisotropy factor in the overdoped region. The differences between these results could be attributed to the uncertainties associated with the finite size of the electrical contacts in the magnetoconductivity measurements. Other measurements of $\gamma$ in Ba(Fe$_{1-x}$Ni$_x$)$_2$As$_2$ in the literature focused in the optimal doping level \cite{niPRB10,tao,sun,shahbazi}. The values found in these works ($\gamma\approx1.7-3$) are close to the one observed in our optimally-doped crystal, further confirming the reliability of our analysis.

\begin{figure}[t]
\begin{center}
\includegraphics[scale=.5]{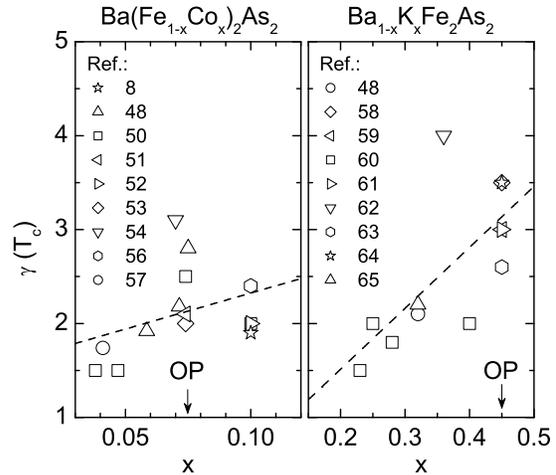}
\caption{Anisotropy factor (near $T_c$) against the doping level in the most studied compounds of the 122 family, according to data in the literature. The optimal doping for each compound is indicated by an arrow. The lines are linear fits.}
\label{figure6}
\end{center}
\end{figure}

In other compounds of the 122 family, in particular in the electron-doped Ba(Fe$_{1-x}$Co$_x$)$_2$As$_2$ and in the hole-doped Ba$_{1-x}$K$_x$Fe$_2$As$_2$, there are some studies including non optimally-doped samples \cite{putti,sun,niPRB08,tanatarPRB09,yamamoto01,prozorov,nojima,maiorov,hanisch,Vinod,altarawneh,niPRB08_2,wang,martin,kim09,welpPRB09,jiao,gasparov}. The corresponding $\gamma$ values are plotted against $x$ in Fig.~\ref{figure6}. These values correspond to temperatures close to $T_c$, where no appreciable differences were observed between $\gamma_\lambda\equiv \lambda_c/\lambda_{ab}$ and $\gamma\equiv\xi_{ab}/\xi_c$, and may be directly compared to our present results. In spite of the dispersion, it seems that $\gamma(T_c)$ also tends to increase with the doping level, although the large values observed here are not observed. It is worth noting, however, that in these works the maximum studied doping level ($x_{\rm max}$) relative to the optimal one ($x_{\rm op}$, indicated as arrows in the figures) are below the one reached in our work: in Ba$_{1-x}$K$_x$Fe$_2$As$_2$ there are no data in the overdoped region, in Ba(Fe$_{1-x}$Co$_x$)$_2$As$_2$ $x_{max}/x_{op}=1.3$, while in our present work $x_{max}/x_{op}=2$.

Just for completeness, in Table 2 we summarize the anisotropy factors found in other families of Fe-based superconductors. For these compounds no clear dependence of the anisotropy factor with the doping level is observed. 1111 compounds are more anisotropic than the ones from the 122 family (the anisotropy factor at optimal doping is about $\sim5$), and they even present 2D characteristics. In spite of that, values of the anisotropy factor as large as the ones observed here in highly overdoped Ba(Fe$_{1-x}$Ni$_x$)$_2$As$_2$ were still not reported.

\begin{table*}[t]
\begin{center}
\begin{tabular}{llcccc}
\hline
Family & Compound & $x$ & $T_c$(K)  & $\gamma(T_c)$ & Ref. \\ 
\hline  
1111 & NdFeAsO$_{1-x}$F$_x$ & 0.18 & 47  & 4 & \cite{welpPRB08}\\
& & 0.18 & 52  & 4.5 & \cite{jia08}\\
& & 0.18 & 46 & 3.9 & \cite{wangAdvMater09}\\
& & 0.3 & 47.4 & 5 & \cite{putti} \\ 
& & 0.35 & 40 & 3.5 & \cite{eisterer}\\
& SmFeAsO$_{1-x}$F$_x$ & 0.15 & 49.5 & 8 & \cite{welpPRB11} \\
& & 0.2 & 42 & 6.5 & \cite{lee09}\\
& & 0.2 & 52.3 & 2 & \cite{prando}\\
& SmFeAsO$_{0.7}$F$_{0.25}$ &   & 49 & 7.5 & \cite{karpinski} \\
& PrFeAsO$_{1-x}$ & 0.1  & 34 & 4 & \cite{okazaki09}\\
& & 0.3 & 45 & 5 & \cite{shirage09}\\
& SmFeAsO$_{1-x}$ & 0.15 & 50.5 & 5 & \cite{lee09}\\
\hline
111 & LiFeAs	&		&	17.6	&	2.5	&	\cite{cho}\\
\hline
11 & FeSe$_{1-x}$Te$_x$ 	&	0.5	&	14.6	&	1.6	&	\cite{bendele}\\
& & 0.5 & 14.5 & 1.1-1.9 & \cite{putti} \\
& Fe(Se,Te)	&	 	&	13.6	&	3	&	\cite{kazumasa}\\
\hline
\end{tabular}
\end{center}
\caption{Summary of values in the literature for the anisotropy factor near $T_c$ in compounds of the 1111, 111, and 11 families.}
\label{tabla3}
\end{table*}

\section{Conclusions}
We have presented measurements of the magnetic susceptibility just above the superconducting transition of the iron pnictide Ba(Fe$_{1-x}$Ni$_x$)$_2$As$_2$ with different doping levels. The measurements were performed with magnetic fields up to 7~T applied both parallel and perpendicular to the FeAs ($ab$) layers. The excellent structural and stoichiometric quality of the crystals studied, which show sharp diamagnetic transitions, allowed to obtain accurate data of the fluctuation effects in a wide temperature range above the superconducting transition. These experimental results were analyzed in terms of a Gaussian Ginzburg-Landau approach for 3D anisotropic superconductors valid in the finite-field (or Prange) fluctuation regime. This single-band approach was found to be in excellent agreement with the data up to the highest reduced temperatures and magnetic fields explored. This suggests that the coherence lengths in the different bands of this superconductor are not very different, or that the field scale at which multiband effects are expected to be observable is not in the field range of our experiments. The analysis allowed to determine the dependence of the in-plane $\xi_{ab}(0)$ and transverse $\xi_c(0)$ coherence lengths with the doping level. The anisotropy factor, $\gamma=\xi_{ab}(0)/\xi_c(0)$ was found to increase from $\sim3$ at optimal doping ($x=0.05$) to $\sim15$ well inside the overdoped region ($x=0.10$). These results provide a quantitative confirmation of the conclusions proposed for the same compounds in Ref.~\cite{reySST13} from measurements of the fluctuation-induced magnetoconductivity. It would be desirable to check whether such a large increase of the anisotropy factor is also present in other Fe-based superconductors at high doping levels.

\ack

This work was supported by the Xunta de Galicia (grant no. GPC2014/038). AR-A acknowledge financial support from Spain's MICINN through a FPI grant (no. BES-2011-046820). The work at IOP, CAS in China is supported by NSFC Program (No. 11374011), MOST of China (973 project: 2011CBA00110) and CAS (SPRP-B: XDB07020300).

\end{document}